\documentclass[11pt]{article}
\usepackage{url} 
\usepackage{lineno}
\usepackage{soul}
\usepackage{amsmath}
\usepackage{amssymb}
\usepackage{booktabs}

\usepackage{natbib}
\usepackage{graphicx}
\usepackage{authblk}

\usepackage[
  a4paper,
  left=2cm,
  right=2cm,
  top=2.5cm,
  bottom=3cm
]{geometry}

\newcommand{\dy}{\, \mathrm{d}y} 
\newcommand{\dx}{\, \mathrm{d}x} 
\newcommand{\eps}{\varepsilon} 
\newcommand{\thalf}{t_\text{50}}

\renewenvironment{abstract}
  {\begin{center}
     \bfseries Abstract
   \end{center}
   \begin{center}
   \begin{minipage}{0.85\textwidth}
   }
  {\end{minipage}
  \end{center}}

\title{Universal ordinary differential equations and the \linebreak parameterization of warm-rain processes}

\author{
  Axel Seifert\thanks{Corresponding author: axel.seifert@dwd.de} \\
  \vspace{-3mm}
  {\small Deutscher Wetterdienst, Offenbach, Germany}
}

\begin{document}

\maketitle

\begin{abstract}
  The parameterization of warm-rain formation is a long-standing
  problem in cloud microphysics because the evolution of bulk cloud
  variables cannot be derived uniquely from the kinetic collection
  equation (KCE). Universal ordinary differential equations (UODEs)
  provide a scientific machine learning framework that combines neural
  networks with ordinary differential equations and can learn
  dynamical operators directly from time series. Here, the UODE
  framework is applied to warm-rain formation by training on
  trajectories generated with a super-droplet model that approximates the
  KCE. In contrast to most previous machine-learning approaches, only the
  prognostic state variables are used for training, without requiring
  autoconversion or accretion rates as targets. The learned closure
  accurately reproduces the KCE trajectories within the training
  domain and, although not explicitly constrained to do so, yields
  autoconversion and accretion rates that closely resemble those
  diagnosed from the KCE. Analyzing the learned operator provides
  new insight into the structure of the widely used Seifert and Beheng
  (2001) warm-rain parameterization. In particular, it suggests that
  the empirical suppression of accretion primarily compensates for an
  overestimation of autoconversion at small rain fractions. Motivated
  by these findings, a refined analytical formulation is proposed that
  improves agreement with the KCE while retaining the simplicity of
  the original parameterization. The results demonstrate that UODEs
  can serve not only as accurate surrogate models but also as a tool
  for understanding and improving analytical parameterizations of
  cloud microphysical processes.
\end{abstract}

\newpage

\section{Introduction}

Parameterizing warm-rain formation within a grid volume of a weather
or climate model is a classic problem in cloud physics.  In general,
this includes assumptions about sub-grid variability of water vapor
and liquid water within such a grid volume. For resolved processes,
though, the warm-rain parameterization problem can be formulated as
the task of finding a finite set of ordinary differential
equations (ODEs) that reproduce the evolution of cloud and rain water as
described by the kinetic collection equation (KCE). The latter is also
known as the Smoluchowski equation. In such a two-category bulk scheme
the mass transfer from cloud water to rain due to the interaction of
cloud droplets is called autoconversion, whereas mass transfer due to
collection of cloud droplets by raindrops is called accretion. 

Autoconversion is highly non-linear and represents the more difficult
part of the warm-rain closure problem. One of the first
parameterizations of autoconversion based on the KCE was proposed by
\citet{Berry-Reinhardt-1974d}, and others followed
\citep{Lüpkes-1989,Beheng-1994,Seifert-Beheng-2001,Sant-2013,Lee-Baik-2017}.
In contrast, some schemes address the broader warm-rain
parameterization problem by explicitly or implicitly representing
subgrid variability of the cloud field
\cite[][e.g.]{Khairoutdinov-Kogan-2000,Larson-Griffin-2013}, whereas
other schemes, following the tradition of \citet{Kessler-1969}, are
not explicit about the specific problem they solve
\citep{Tripoli-Cotton-1980,Liu-2004a,Liu-2004b,Liu-2006}.  The
difficulties associated with autoconversion have in recent years
renewed interest in single-category warm-rain schemes based on
higher-order moment closures as an alternative formulation
\citep{Igel-2022,Morrison-2025}.  A still informative review of
warm-rain formation and the two-category parameterization problem is
provided by \citet{Beheng-2010}.

The parameterization of \citet[SB2001 hereafter]{Seifert-Beheng-2001},
which serves as a benchmark in the present study, has been implemented
in numerous numerical weather prediction and large-eddy simulation
models, including COSMO, ICON, NICAM, PALM, MicroHH, DALES, MIMICA,
UCLA-LES, and NU-WRF
\citep{Baldauf-2011,Zangl-2015,Seiki-2022,Maronga-2015,Anurose-2020,Heus-2010,Savre-2014,Stevens-Seifert-2008,Stanford-2025}.
It has also been employed in the climate configurations ICON-XPP and
ICON-Sapphire and evaluated in experimental versions of CAM
\citep{Hohenegger-2023,Muller-2025,Bogenschutz-2018,Gettelman-2020}.
Because SB2001 is derived
solely from the KCE, it does by itself not account for sub-grid
variability. Moreover, its formulation does not readily permit the
inclusion of such effects. Consequently, it is most appropriate for
well-resolved simulations, particularly LES, and, without additional
efforts, less suitable for coarse-resolution climate models.

Recent advances in scientific machine learning allow to
incorporate flexible neural-network representations in ODE systems while retaining
their dynamical structure. In this work, the universal ordinary
differential equation (UODE) framework first introduced by
\citet{Rackauckas-2020} is used to learn representations of
autoconversion and accretion directly from solutions of the KCE.  The
current paper is following up on \citet[SR20
  hereafter]{Seifert-Rasp-2020}, who made a first attempt to machine
learn warm-rain processes using a simpler supervised learning
approach.  Other machine learning approaches to warm-rain processes
were \citet{Gettelman-2021}, \citet{Alfonso-Zamora-2021} and more
recently \citet{Sharma-Greenberg-2025} and \citet{deJong-2026}.

Unlike supervised learning approaches that require process rates as
targets, the UODE framework is trained using only time series of the
state variables, here liquid water content and droplet number
concentration, rather than process rates. Although only trained on
state variables, the learned model represents the warm-rain
differential operator, i.e., the autoconversion and accretion
rates. In other words, the UODE learns the closure itself.

The goal of the current study is not to propose another neural network
based scheme. Rather, the UODE framework is used as a tool for
understanding warm-rain processes and for guiding the development of
improved analytical parameterizations. Existing analytical
parameterizations necessarily introduce closure assumptions whose
impact and realism are difficult to assess. A UODE trained directly on
KCE solutions provides a flexible reference model against which
analytical parameterizations can be evaluated and from which improved
closure formulations may be inferred.

The paper is organized as follows. In section 2 the KCE and the
corresponding warm-rain equations are introduced including SB2001. In
section 3 the UODE model is explained, and in section 4 the training
of the UODE is described. In section 5 the resulting neural network
representation is analyzed and the compared with the KCE benchmark and
SB2001. Based on these insights, a new parameterization is suggested
in section 6 and compared with SB2001. The paper ends with some
conclusions in section 7.

\section{Warm-rain equations}

The evolution of the droplet size distribution due to collision-coalescence is described by the
kinetic collection equation (KCE)
\begin{linenomath*}
  \begin{equation}
    \label{eq:sce}
    \frac{d f(x)}{dt} = \frac{1}{2} \int_0^x f(x-y) f(y) K(x-y,y) \dy
    - \int_0^\infty f(x) f(y) K(x,y) \dy
  \end{equation}
\end{linenomath*}
also known as the Smoluchowski equation or the quasi-stochastic collection
equation. Here
$f(x)$ is the number of droplets per unit volume in the mass range
$[x,x+\dx]$, and $x$ and $y$ are the drop mass. The
collision-coalescence kernel $K(x,y)$ describes the physics of the
collision process and is specified as
\begin{linenomath*}
  \begin{equation}
    \label{eq:kernel}
    K(x,y) = {\pi} \, [ r(x) + r(y) ]^2 \, |v(x)-v(y)| \, E_{coll}(x,y)
  \end{equation}
\end{linenomath*}
where $r(x)$ is the drop radius, $v(x)$ is the terminal fall velocity
and $E_{coll}(x,y)$ is the collision efficiency. In the following, the
collision efficiency of \citet{Hall-1980} is used with modifications
by \citet{Beheng-1994}. The KCE has a corresponding ODE system for the
moments
\begin{linenomath*}
\begin{alignat}{2}
  &M^{(k)} &&= \int_0^\infty x^k f(x) \dx          \\
\end{alignat}
\end{linenomath*}
with
\begin{linenomath*}
  \begin{equation}
    \label{eq:sce-moments}
    \frac{d M^{(k)} }{dt} = \frac{1}{2} \int_0^\infty \int_0^\infty f(x) f(y) K(x,y)
          \left [ (x+y)^k - x^k - y^k \right ] \dx \dy
  \end{equation}
\end{linenomath*}
which is fully equivalent to the KCE \citep[e.g.][]{Drake-1972}. For a
non-linear kernel $K$ this constitutes an infinite dimensional system
and, hence, a closure problem for any finite number of predicted
moments.

A two-category bulk warm-rain scheme predicts only partial moments of
the distribution $f(x,t)$ for cloud and rain categories defined as:
\begin{linenomath*}
\begin{alignat}{2}
  &M_c^{(k)} &&= \int_0^{x_*} x^k f(x) \dx          \\
  &M_r^{(k)} &&= \int_{x_*}^\infty x^k f(x) \dx
\end{alignat}
\end{linenomath*}
and drops smaller than the mass $x_*=2.6\times 10^{-10}$ kg are
classified as cloud droplets, whereas larger droplets are classified
as raindrops. $M_i^{(0)}=N_i$ is the number density and
$M_i^{(1)}=L_i$ the mass density or cloud resp.~rain water content.

The evolution of the partial moments can be written as a system of
ordinary differential equations (ODEs) for the partial moments $N_c$,
$L_c$, $N_r$ and $L_r$ given by
\begin{linenomath*}
\begin{alignat}{3}
  &\frac{dL_c}{dt} &&= - AU - AC &&  \label{eq:ode1} \\
  &\frac{dL_r}{dt} &&= + AU + AC &&  \label{eq:ode2}\\
  &\frac{dN_c}{dt} &&= - 2 AU_N - AC_N - SC_c 
                   &&= - \frac{2}{x_*} AU - \frac{1}{\bar{x}_c} AC - SC_c \label{eq:ode3} \\
  &\frac{dN_r}{dt} &&= + AU_N - SC_r &&= + \frac{1}{x_*} AU - SC_r \label{eq:ode4}
\end{alignat}
\end{linenomath*}
with the mean cloud droplet mass $\bar{x}_c = L_c/N_c$. The
autoconversion rate $AU$, the accretion rate $AC$ and the two
self-collection rates $SC_c$ and $SC_r$ are unknown. Specifying these
process rates in terms of the predicted partial moments constitutes
the warm-rain parameterization (or closure) problem. Here the
assumptions have been used that autoconversion creates droplets of the
mass $x_*$ and accretion collects on average the cloud droplets with
mass $\bar{x}_c$ to replace the number rates of autoconversion and
accretion, $AU_N$ and $AC_N$ \citep{Beheng-1994,Beheng-2010}.

Among the many proposed closures, the formulation of SB2001 has become
one of the most widely used KCE-based parameterizations and serves as
the reference model in the present study.
\begin{linenomath*}
\begin{alignat}{2}
  &AU_{sb} &&= \frac{k_{cc}}{20 x_*} \, \frac{(\nu+2)(\nu+4)}{(\nu+1)^2} \, L_c^2 \, \bar{x}_c^2 \, \Phi_{au}(\tau)\\
  &AC_{sb} &&= k_{cr} \, L_c L_r \, \Phi_{ac}(\tau)
\end{alignat}
\end{linenomath*}
where $k_{cc} = 9.44 \times 10^9$ m$^3$ kg$^{-2}$ $s^{-1}$ and $k_{cr} =
5.78$ m$^3$ kg$^{-1}$ $s^{-1}$ are coefficients of the Long kernel
\citep{Long-1974} and $\nu$ is the shape parameter of an assumed Gamma
distribution. The functions $\Phi_{au}(\tau)$ and $\Phi_{ac}(\tau)$ are given by
\begin{linenomath*}
\begin{alignat}{2}
  &\Phi_{au}(\tau) &&= 1 + \frac{600 \tau^{0.68} (1 - \tau^{0.68})^3}{(1-\tau)^2} \label{eq:phi_au} \\
  &\Phi_{ac}(\tau) &&= \left ( \frac{\tau}{\tau + 5 \times 10^{-4} } \right )^4 \label{eq:phi_ac}
\end{alignat}
\end{linenomath*}
and depend only on the rain fraction $\tau=L_r/(L_c+L_r)$.  For
convenience, $\Phi_{au}$ is written here in a form that differs from
the definition used in SB2001 and SR20. The resulting autoconversion
rate is mathematically equivalent to their formulation. The
corresponding cloud and rain self-collection rates are
\begin{linenomath*}
\begin{alignat}{2}
  &SC_{c} &&= k_{cc} \, \frac{\nu+2}{\nu+1} \, L_c^2 \\
  &SC_{r} &&= k_{rr} \, L_r N_r
\end{alignat}
\end{linenomath*}
with $k_{rr} = 4.33$ m$^3$ kg$^{-1}$ $s^{-1}$.

\section{The UODE model}

The UODE model replaces the analytical closures for autoconversion and
accretion by a neural network representation, while retaining the
analytical formulation of the remaining terms:
\begin{linenomath*}
\begin{alignat}{2}
  &\frac{dL_c}{dt} &&= - AU_\text{nn} - AC_\text{nn}   \label{eq:uode1} \\
  &\frac{dL_r}{dt} &&= + AU_\text{nn} + AC_\text{nn}  \label{eq:uode2}\\
  &\frac{dN_c}{dt} &&=  - \frac{2}{x_*} AU_\text{nn} - \frac{1}{\bar{x}_c} AC_\text{nn} - SC_c \label{eq:uode3} \\
  &\frac{dN_r}{dt} &&= + \frac{1}{x_*} AU_\text{nn} - SC_r \label{eq:uode4}
\end{alignat}
\end{linenomath*}
while keeping the analytical parameterizations of SB2001 for the
selfcollection terms. For the neural network the prognostic variables are
combined in a state vector
\begin{linenomath*}
  \begin{equation}
    \mathbf{u} = \begin{pmatrix} L_c \\ L_r \\ N_c \\ N_r \end{pmatrix}
  \end{equation}
\end{linenomath*}
and the input for the neural network is a log-transform 
\begin{linenomath*}
  \begin{equation}
    \mathbf{x} = \log ( \max( \mathbf{u}, \boldsymbol{\epsilon}) ).
  \end{equation}
\end{linenomath*}
The threshold vector is $\boldsymbol{\epsilon} = (\epsilon_L,
\epsilon_L, \epsilon_N, \epsilon_N)^T$ with $\epsilon_L=10^{-12}$ and
$\epsilon_N=10^{-1}$. Here the $\max$ operator of two vectors is interpreted elementwise.

The transformed input vector is mapped to the output vector
\begin{linenomath*}
  \begin{equation}
    \mathbf{y}=\mathcal N_\theta(\mathbf{x}),
  \end{equation}
\end{linenomath*}
where $\mathcal N_\theta:\mathbb R^4\rightarrow\mathbb R^2$ is a
fully connected feed-forward neural network with architecture
$4\rightarrow64\rightarrow64\rightarrow32\rightarrow2$. The
hidden layers use the hyperbolic tangent activation function and the
output layer is linear. The output vector $\mathbf{y}$ is then
transformed back to the physical variables and yields the desired process
rates:
\begin{linenomath*}
  \begin{equation}
    \begin{pmatrix} AU_\text{nn} \\ AC_\text{nn} \end{pmatrix}
    = \exp( \mathbf{y} )
  \end{equation}
\end{linenomath*}
The logarithmic transformation serves two purposes. First, it
compresses the dynamic range of the state variables, which span many
orders of magnitude. Second, together with the exponential output
transformation, it guarantees positive autoconversion and accretion
rates.

The parameters $\theta$ of the neural network are determined by minimizing
a loss function that measures the mismatch between simulated and
reference trajectories, as described in the following section.

\section{Training of the UODE model}

As training data, highly accurate numerical solutions of the kinetic
collection equation (KCE) are generated using the super-droplet method
of \citet{Shima-2009}. Owing to the Monte Carlo formulation,
instantaneous process rates exhibit sampling noise, but only the
prognostic state variables are used for training the UODE. Note that
the Monte-Carlo algorithm implements a more general equation which
takes into account stochastic fluctuations
\citep{Dziekan-Pawlowska-2017}. The KCE, Eq.~\eqref{eq:sce}, is the
mean-field approximation of this more general stochastic collection
equation \citep{Alfonso-2008}.

As initial condition, a Gamma distribution for cloud droplets with
\begin{linenomath*}
\begin{equation}
  \label{eq:init}
  f(x) = A x^\nu e^{-Bx}
\end{equation}
\end{linenomath*}
is used where $A$ and $B$ are calculated from the initial mean radius
$\bar{r}_0$ and the initial liquid water content $L_0=L_{c,0}$, and
$\nu$ is the shape parameter, which is set to $\nu=1$ in the current
study.  The training data consist of 24 trajectories with $L_0 \in
[0.5,2]$ g$\,$m$^{-3}$ and $\bar{r}_0 \in [11,20]$ $\mu$m. More than
10,000 super-droplets are used, which ensures a sufficiently converged
solution. Some more details on the super-droplet model are given in
SR20 and also apply to the current study. In contrast to SR20, the
solutions are sampled with a high time resolution of $2\,$s, which
simplifies the training of the UODE.  Although only 24 trajectories
are used, each trajectory samples a different region of the phase space
and contains roughly 1000 time steps, providing approximately 20,000
state-vector increments that constrain the learned differential operator.

Training is performed within the SciML ecosystem of Julia using
OrdinaryDiffEq.jl, SciMLSensitivity.jl, Optimization.jl, and Zygote.jl
\citep{Rackauckas-2020}. The UODE is integrated with an explicit ODE
solver, while adjoint sensitivity analysis is used to compute
gradients of the trajectory cost function with respect to the
neural-network parameters by differentiating through the numerical ODE
solution. These gradients are then passed to the Optimization.jl
framework, which employs the Adam or L-BFGS optimizer to minimize the
cost function.

This approach is analogous to variational data assimilation in
numerical weather prediction, only that for the UODE training the
initial conditions are known exactly and the model is optimized,
whereas in atmospheric data assimilation the model is assumed to be
perfect and the initial condition is estimated. Technically, this is a
very similar procedure to using an adjoint model and gradient descent.

A key ingredient of the training procedure is the definition of the
trajectory loss and the cost function. Consistent with the logarithmic
transformation of the network variables, the state loss is defined as
\begin{linenomath*}
  \begin{equation}
    \mathcal{L}
    \!\left(  \mathbf{u}, \mathbf{u}^{\mathrm{KCE}} \right )
    =
    \sum_{i=1}^{4} w_i
    \left [ \log\!\left(\max(u_i,\eps_i) \right)
          - \log\!\left(\max(u_i^{\mathrm{KCE}},\eps_i)\right) \right]^2,
  \end{equation}
\end{linenomath*}
with $\eps_1=\eps_2=10^{-9}$, $\eps_3=\eps_4=1$, $w_1=w_2=1$,
and $w_3=w_4=0.1$. The trajectory loss is obtained by averaging this loss over all time steps $K$:
\begin{linenomath*}
\begin{equation}
  \mathcal{J} = \frac{1}{K} \sum_{k=1}^{K}
           \mathcal{L} \!\left( \mathbf{u}_k, \mathbf{u}_k^{\mathrm{KCE}} \right).
\end{equation}
\end{linenomath*}
During optimization, $\mathcal{J}$ is further averaged over the
trajectories contained in each mini-batch. This averaged mini-batch
loss is the cost function, which is minimized by the optimizer. To
monitor the result after each epoch, the trajectory losses are
recalculated and averaged over all trajectories. This is called the
full trajectory loss in the following.

The weighting emphasizes accurate prediction of the liquid water
variables, which are the primary quantities of interest for warm-rain
evolution. Larger errors are permitted in the number densities, both
because their governing equations involve additional approximations
that are outside the scope of the present closure and because they are
of secondary importance for many applications.

The lower bounds for $L_c$ and $L_r$ in the loss function,
$\eps_1=\eps_2=10^{-9}$, are chosen substantially larger than the
lower bound $\epsilon_L$ used in the logarithmic transformation of the
network inputs. This distinction is important. Values below the loss
threshold correspond to an effective no-rain state and whether $L_r$
is $10^{-9}$ or $10^{-10}$ seems unimportant from a physical point of
view.  Ignoring errors in this regime prevents the optimization from
being dominated by insignificant fluctuations in extremely small
rainwater contents.  Moreover, the interval between the network lower
bound $\eps_L$ and the loss threshold $\epsilon_2$ provides several
orders of magnitude in which the rainwater variable may evolve without
contributing to the loss.  As shown in the next section, the UODE
exploits this unconstrained region to establish an effective closure.
Thus, although these tiny values of $L_r$ do not contribute directly
to the cost function, they still influence the learned dynamics.

Both, the weighting of the loss function and the $\eps_i$ thresholds,
reflect the fact that a two-category two-moment representation cannot
be expected to reproduce all aspects of the KCE evolution exactly. The
closure problem is intrinsically ill-posed because the full particle
size distribution contains more information than the four prognostic
variables retained in the bulk representation. Consequently, the UODE
is not required to reproduce every component of the state vector with
identical accuracy. These choices provide the flexibility required for
the UODE to find an approximate closure; enforcing an exact
reproduction of all KCE variables over the full trajectory would
over-constrain the problem and limit convergence.

\begin{figure}[t]
  \noindent\includegraphics[width=\textwidth]{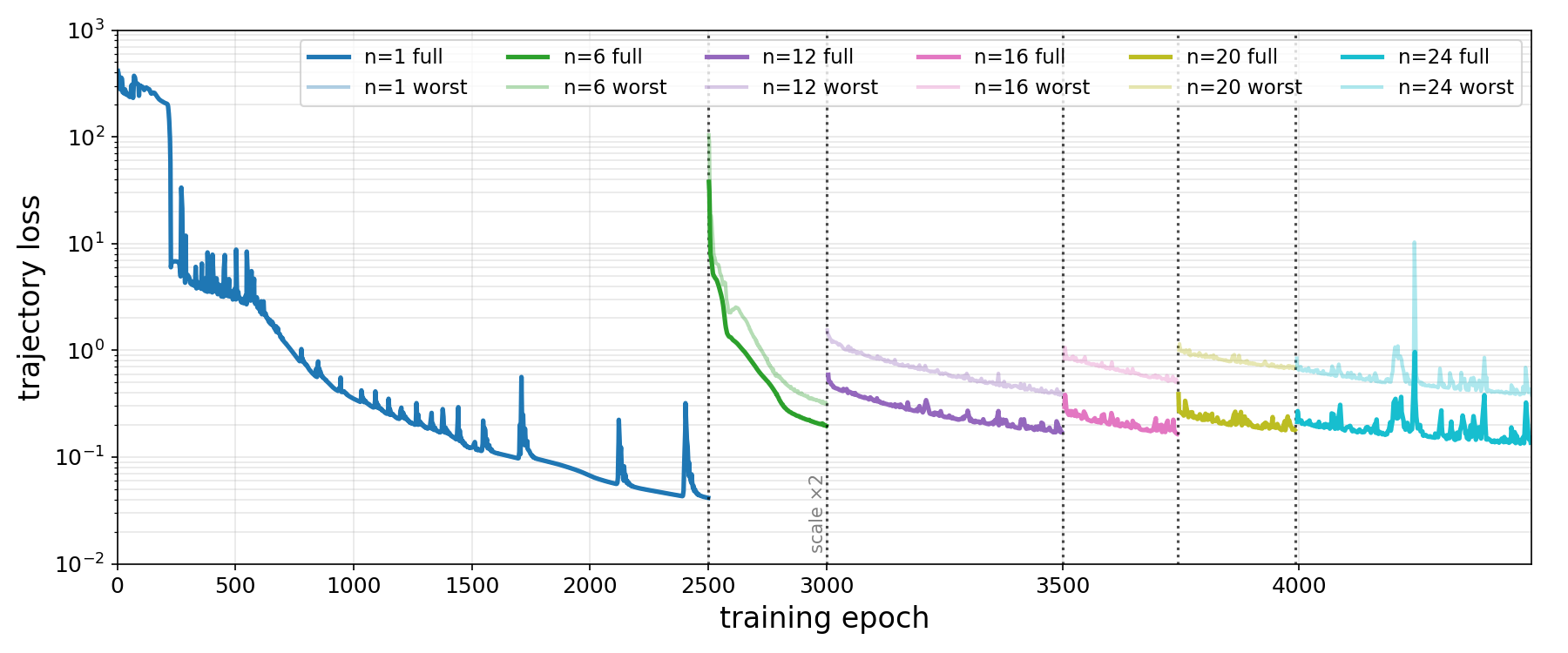}
  \vspace{-8mm}
  \caption{\label{Fig:optimization_curve} Optimization curve for the
    continuation training procedure showing the full trajectory loss
    and the worst trajectory loss as functions of the training
    epoch. The worst trajectory loss denotes the loss of the least
    accurately reproduced trajectory within the current training
    set. The legend indicates the number of trajectories $n$ included
    in the current training set. For clarity, the x-axis is compressed
    over the first 3000 training epochs.}
\end{figure}

To make repeated differentiation through long trajectories
computationally feasible, the UODE is integrated using an explicit
solver with a constant time step of 2$\,$s. Although implicit solvers are
attractive for stiff systems, differentiating through them with
automatic differentiation remains expensive for the
thousands of optimization epochs required for operator
learning. Training employs mini-batch Adam optimization with batches
of four to six trajectories and gradient clipping as a pragmatic
stabilization strategy for the explicit integration.

The training strategy follows a continuation (curriculum) learning
approach. Instead of exposing the network to the full parameter space
from the beginning, training starts with a single KCE trajectory,
allowing the network to capture the basic dynamics of the
collision–coalescence operator. Additional KCE trajectories are then
introduced progressively, expanding the training set from one to six,
twelve, sixteen, and beyond. Each new trajectory explores a different
region of phase space characterized by distinct combinations of cloud
water, rain water, cloud droplet and raindrop number density, and
their associated timescales. Some previously learned trajectories are
retained throughout training so that the network continually expands
its domain of validity while maintaining skill on regions of parameter
space that have already been learned, thereby reducing the risk of
catastrophic forgetting. Figure~\ref{Fig:optimization_curve} shows the
optimization curve for the complete continuation training procedure.

For the initial training of the first trajectory, a learning rate of
$5\times10^{-3}$ is used for the Adam optimizer. During continuation
learning, the learning rate is reduced to $5 \times 10^{-4}$.
Although this relatively high learning rate introduces noticeable
fluctuations during optimization, it is more effective in reducing the
loss than lower learning rates. The resulting stochasticity
helps the Adam optimizer explore the loss landscape and escape local
minima. Each subset of trajectories is trained until the full
trajectory loss drops below 0.1. For the final training set comprising
24 trajectories, the full trajectory loss at the end of the training
is 0.1334.

\section{Results}

\subsection{Trajectory-level performance}

The results are evaluated using a testing set of five KCE trajectories
that were never used during training. Table 1 summarizes the initial
conditions together with three performance metrics: the trajectory
loss $\mathcal{J}$ used for training, an additional quadratic trajectory
loss $\mathcal{J}_\text{2}$, and the characteristic conversion time
$\thalf$, defined as the time at which 50 \% of the initial cloud
water has been converted into rainwater.

The weighted quadratic loss in physical space is defined as
\begin{linenomath*}
  \begin{equation}
    \mathcal{L}_\text{2}
    \!\left(  \mathbf{u}, \mathbf{u}^{\mathrm{KCE}} \right )
    =
    \sum_{i=1}^{4} 
    \left [ s_i \left( u_i - u_i^{\mathrm{KCE}} \right) \right]^2,
  \end{equation}
\end{linenomath*}
with $s_1=s_2=10^{4}\,$kg$^{-1}\,$m$^{3}$ and
$s_3=s_4=10^{-8}\,$m$^{3}$. As for the logarithmic loss, the quadratic
loss is averaged over the full trajectory to obtain
$\mathcal{J}_\text{2}$. In addition to Table 1, Figure
\ref{Fig:timeseries_states} visualizes the time series of the state
variables for all 5 trajectories of the testing set for the KCE, the
UODE, SB2001, and a new parameterization that will be introduced in
detail in the next section.

The time series in Figure \ref{Fig:timeseries_states} show the typical
evolution of warm-rain formation. In the earliest stage, very little
happens and only $N_c$ decreases due to self-collection. At some
point, warm-rain formation becomes increasingly efficient. Once a
sufficient number of raindrops has formed through autoconversion,
accretion becomes dominant, leading to a rapid decrease in $L_c$ and a
corresponding increase in $L_r$. Soon after, all cloud water is
converted to rain. All models agree reasonably well with the KCE
reference. The time series of $N_r$ are particularly interesting,
because $N_r$ is only created by autoconversion and later depleted by
self-collection of raindrops, leading to a pronounced maximum during
the transition from the cloud-dominated to the rain-dominated state.
Therefore, $N_r$ provides an additional constraint on the
autoconversion process and serves as a sensitive diagnostic of how
well the UODE has learned this part of the closure. The UODE
underestimates $N_r$ for all trajectories except trajectory 4, for
which it shows a clear overestimation. The analytical
parameterizations show no comparable outlier in $N_r$, suggesting
again that the UODE has not fully captured the dependence of the
closure on $L_c$ and $\bar{r}_0$. The initial decrease in $N_c$ is
slightly too strong for the UODE and both analytical
parameterizations. This suggests that $k_{cc}$ is somewhat too large
for self-collection. The UODE cannot improve this discrepancy because
self-collection has not been replaced by a neural network
approximation.

Table 1 and Figure \ref{Fig:timeseries_states} show that the UODE
provides a very good approximation for trajectories 1--4, but is
slightly less accurate for trajectory 5 with $L_0=0.3$
g$\,$m$^{-3}$. This is easily explained by the fact that this value of
$L_0$ is outside the training data, which span only
$[0.5,2]$~g$\,$m$^{-3}$. Although this represents a relatively simple
extrapolation, the UODE loses skill for small $L_0$ despite the known
scaling behavior of the KCE. This reveals that the UODE has not fully
recovered the underlying physical scaling, but rather learned an
accurate approximation within the range covered by the training
data. This it does very well, though.  The SB2001 scheme performs
well, especially in $\thalf$, less so for the quadratic loss, and is
clearly worse in the logarithmic loss.  This suggests that SB2001
primarily accumulates relative errors during the early stages of rain
formation, when rainwater contents remain small. The new
parameterization is comparable in $\thalf$, but closer to the
UODE in both losses than SB2001. The new parameterization generalizes
particularly well to trajectory 5 with low cloud water content and
performs slightly better than both the UODE and SB2001 for this case.

\begin{table}
\centering
\caption{Comparison of the KCE, UODE, SB2001, and the new parameterization
for the five trajectories of the testing set. $L_0$ denotes the initial cloud liquid water
content (g\,m$^{-3}$), and $\bar{r}_0$ the initial mean cloud droplet radius
($\mu$m). The first block shows the characteristic time $t_{50}$ (min),
defined as the time at which 50\% of the initial cloud water has been
converted to rain. The second and third blocks show the logarithmic and
the quadratic trajectory losses relative to the KCE solution.}
\label{tab:comparison}
\vspace{11pt}
\begin{tabular}{ccrrrrr}
\toprule
Trajectory & $L_0$ & $\bar{r}_0$ & KCE & UODE & SB2001 & New \\
 & (g\,m$^{-3}$) & ($\mu$m) & \multicolumn{4}{c}{$t_{50}$ (min)} \\
\midrule
1 & 1.00 & 14.0 & 20.5 & 20.1 & 19.8 & 20.8 \\
2 & 0.70 & 17.0 & 17.8 & 18.0 & 16.4 & 17.2 \\
3 & 0.70 & 14.0 & 29.0 & 28.6 & 28.3 & 29.7 \\
4 & 0.70 & 12.0 & 43.5 & 43.9 & 43.1 & 42.6 \\
5 & 0.30 & 17.0 & 41.9 & 39.9 & 38.3 & 40.1 \\
\midrule
\multicolumn{3}{l}{} &
\multicolumn{4}{c}{logarithmic trajectory loss $\mathcal{J}$}\\
\midrule
1 & 1.00 & 14.0 & --- & 0.097 & 3.242 & 0.744 \\
2 & 0.70 & 17.0 & --- & 0.085 & 0.777 & 0.434 \\
3 & 0.70 & 14.0 & --- & 0.117 & 2.381 & 0.460 \\
4 & 0.70 & 12.0 & --- & 0.408 & 5.373 & 0.659 \\
5 & 0.30 & 17.0 & --- & 0.945 & 1.421 & 0.838 \\
\midrule
\multicolumn{3}{l}{} &
\multicolumn{4}{c}{quadratic trajectory loss $\mathcal{J}_\text{2}$}\\
\midrule
1 & 1.00 & 14.0 & --- & 0.071 & 0.100 & 0.061 \\
2 & 0.70 & 17.0 & --- & 0.004 & 0.155 & 0.025 \\
3 & 0.70 & 14.0 & --- & 0.017 & 0.028 & 0.052 \\
4 & 0.70 & 12.0 & --- & 0.006 & 0.021 & 0.070 \\
5 & 0.30 & 17.0 & --- & 0.050 & 0.057 & 0.013 \\
\bottomrule
\end{tabular}
\end{table}

\subsection{Process-level performance}

\begin{figure}[t]
  \noindent\includegraphics[width=\textwidth]{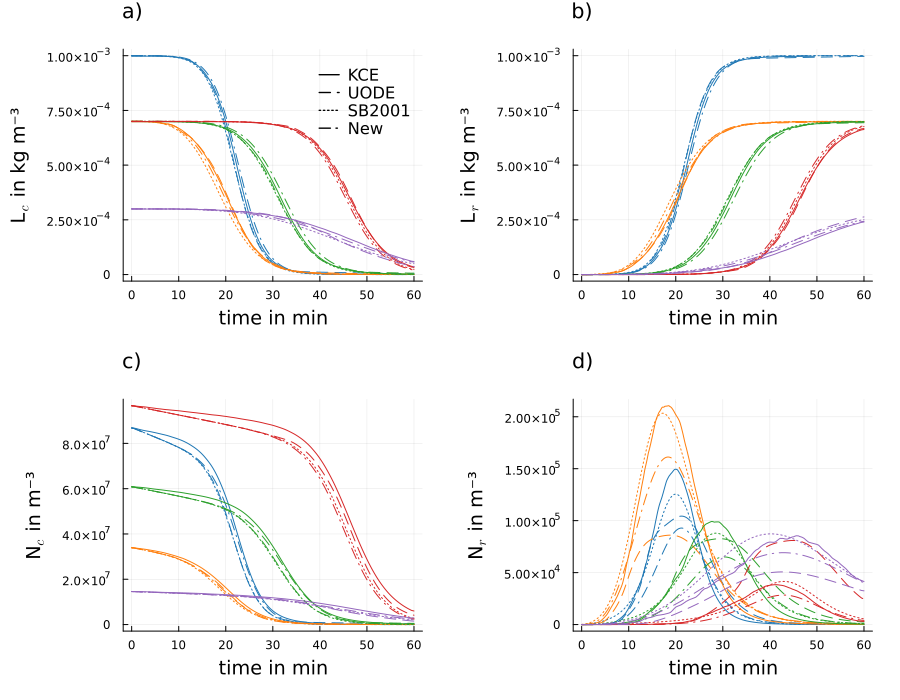}
  \vspace{-8mm}
  \caption{\label{Fig:timeseries_states}
    Time series of the state variables (a) cloud water content
    $L_c$, (b) rain water content $L_r$, (c) cloud droplet number
    density $N_c$, and (d) raindrop number density $N_r$. Shown are
    the five trajectories of the testing set for the KCE (solid), UODE
    (dashed), SB2001 (dotted), and the new parameterization
    (dash-dotted). The color code for the trajectories is 1: blue, 2:
    orange, 3: green, 4: red, 5: magenta using the trajectory numbers
    of Table 1.}
\end{figure}

The behavior of the different closures is better understood by
comparing the actual process rates. The process rates for the KCE
are calculated from the super-droplet approach, see SR20 for
details of the implementation.  Importantly, the process rates shown
here were never used during training. They are reconstructed solely
from fitting the trajectories of the state variables.  The UODE
matches the KCE process rates remarkably well, especially for
trajectory 1 with the fast time evolution. That the UODE has the same
decomposition of AU and AC can be understood by the tendencies of the
number densities.  While the mass equations constrain only the sum
AU+AC; the number equations separate the two processes.  The fact that
$N_r$ has no source term due to accretion and therefore depends
primarily on autoconversion makes the individual process rates
accessible to the UODE method. This also explains why the agreement in
$N_r$ is the least accurate for all schemes in Figure \ref{Fig:timeseries_states}, it
accumulates all errors in the autoconversion rate.  For the slower
trajectory 4 the UODE yields a non-zero autoconversion rate in the
first minutes, which does not match the KCE, but is necessary for any
two-moment bulk scheme as discussed by SR20. Interestingly, this
initial autoconversion rate of UODE is two orders in magnitude smaller
than the corresponding value of SB2001. Hence, the typical
overestimation of the initial autoconversion rate of SB2001 is not
necessary for a successful closure. For accretion all bulk schemes
overestimate the onset of raindrop growth, but match the KCE very well
later on. The new parameterization shows some overestimation of
autoconversion but not as severe as SB2001 and is overall closer to
the KCE and UODE solutions.

\begin{figure}[t]
  \noindent\includegraphics[width=\textwidth]{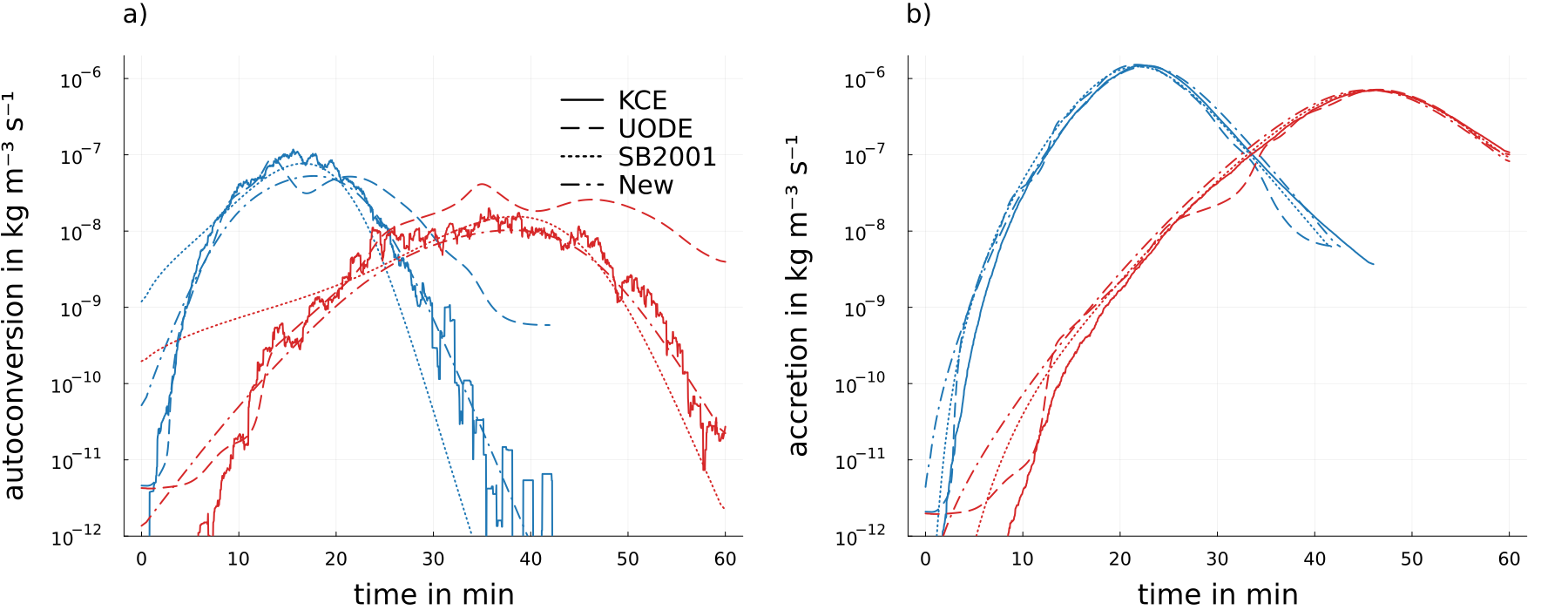}
  \vspace{-8mm}
  \caption{\label{Fig:timeseries_rates} Time series of autoconversion
    and accretion rates for two trajectories of the testing set (1:
    blue, 4: red).  Shown are results for the KCE (solid), UODE
    (dashed), SB2001 (dotted), and the new parameterization
    (dash-dotted). }
\end{figure}

\subsection{What did the UODE learn?}

\begin{figure}[t]
  \noindent\includegraphics[width=\textwidth]{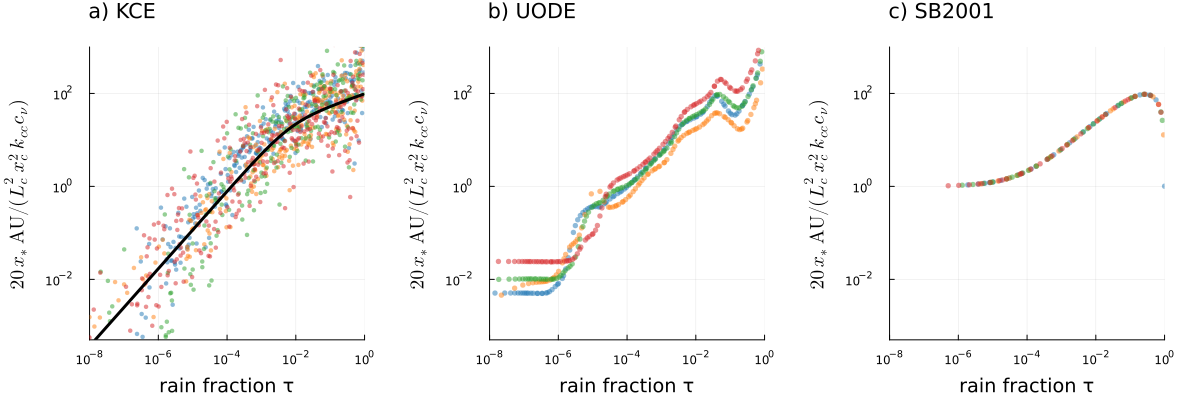}
  \vspace{-8mm}
  \caption{\label{Fig:phiau} Scatter plots of $\Phi_\text{au}(\tau)$
    for the KCE benchmark simulations, and the UODE, and SB2001
    trajectories. Shown are trajectories 1-4 of the test
    set. Different colors represent different trajectories. The factor
    $c_\nu=15/4$ follows from $\nu=1$. The solid black line in (a) is
    a non-linear fit to the KCE data, Eq.~\eqref{kcefit}.}
\end{figure}

Can we understand in some more detail how the UODE warm-rain operator
works? Figure \ref{Fig:phiau} shows the normalized autoconversion rate
using the scaling or prefactor of the SB2001 to nondimensionalize
AU. For the KCE a power-law scaling is evident for small $\tau$ together
with a scale break around $\tau_0=0.007$. Significant noise is
superimposed due to the Monte-Carlo sampling of the super-droplet
model. A nonlinear fit is
\begin{linenomath*}
  \begin{equation} \label{kcefit}
    \Phi_\text{au}^\text{KCE}(\tau) = 26.9 \left ( \frac{\tau}{\tau_0} \right )^{0.83}
                                 \left ( 1 + \frac{\tau}{\tau_0} \right )^{-0.57} 
  \end{equation}
\end{linenomath*}
The UODE reproduces a similar power-law scaling for intermediate
values, approximately for $\tau>10^{-4}$. For very small $\tau$, the
normalized autoconversion rate approaches an approximately constant
value, but the level of this plateau varies slightly between
trajectories. In contrast to the KCE, no clear transition to a
distinct scaling regime is apparent at larger $\tau$.  For SB2001 the
relation simply follows the analytic form of the parameterization,
Eq.~\eqref{eq:phi_au}, and for small $\tau$ it approaches
$\Phi_\text{au}=1$, which does not agree with the KCE benchmark. Two
things are remarkable: First, the UODE has found a broadly similar scaling and
the same decomposition as the SB2001 scheme has, i.e., it has a
$\Phi_\text{au}(\tau)$ function that collapses quite well to a single
functional relationship although some variations between the
trajectories do exist, which means the scaling is only
approximated. Second, the SB2001 scheme does not agree as well with
the KCE benchmark as the UODE. This shows that the assumption of
SB2001 that the limit for small $\tau$ is consistent with the
approximation by the Long kernel, Eq.~(16) of SB2001, is incorrect for
small and even moderate $\bar{r}_0$. This was already evident from
SR20 and is confirmed here. Nevertheless, the result of the Long
kernel does provide a useful guidance for the scaling of the
autoconversion rate.  For the UODE
the autoconversion at large $\tau > 10^{-2}$ is largely unconstrained because
accretion dominates the mass transfer. Hence, the autoconversion rate
can only be estimated from $N_r$, which accumulates all model error
over time making this a difficult inverse problem for large $\tau$.

\begin{figure}[t]
  \noindent\includegraphics[width=\textwidth]{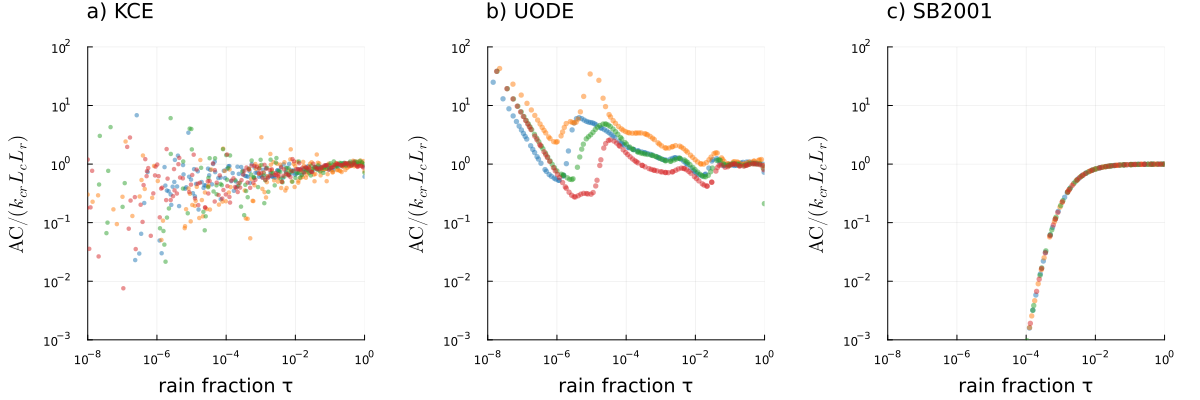}
  \vspace{-8mm}
  \caption{\label{Fig:phiac} Scatter plots of
    $\Phi_\text{ac}(\tau)$ for the KCE benchmark simulations, and the
    UODE, and SB2001 trajectories. Shown are trajectories 1-4 of the
    test set. Different colors represent different trajectories.}
\end{figure}

Figure \ref{Fig:phiac} shows a similar scatter plot for normalized
accretion or $\Phi_\text{ac}(\tau)$. Here the result is different. The
KCE data fluctuates around one suggesting $\Phi_\text{ac}(\tau) \equiv
1$, the UODE overestimates $\Phi_\text{ac}(\tau)$ especially for small
$\tau$, whereas the prescribed functional dependency of SB2001 drops
off for small $\tau$ and artificially suppresses accretion.  This kind
of suggests that both, UODE and SB2001, use the accretion for
fine-tuning the time evolution of rain water content. Both do not
strictly follow the data as provided by the KCE benchmark. Of course,
neither the UODE nor SB2001 have seen this process-level data.

Overall, the UODE reconstructs an accurate approximation to the
warm-rain operator using only trajectories of the four prognostic
state variables. It is not obvious a priori that trajectories of the
state variables alone contain sufficient information to tackle
this nonlinear inverse problem.  Moreover, the closure problem is
inherently ill-posed. As shown by SR20, no two-moment closure can
exactly reproduce the KCE dynamics. This limitation is particularly
evident for small initial radii $\bar{r}_0$, making the learning
problem especially challenging. Given these difficulties, the accuracy
of the learned UODE closure is remarkable.

\section{A new warm-rain parameterization}

The results of the previous section suggest that a reformulation of
the SB2001 warm-rain scheme is possible. First, the super-droplet
benchmark simulations indicate that a correction factor for accretion
as a function of $\tau$ is not required. Second, the UODE closure
shows that the large autoconversion rate predicted by SB2001 for small
$\tau$ is unnecessary and that substantially smaller values are more
consistent with the KCE.

This motivates replacing Eqs.~\eqref{eq:phi_au} and \eqref{eq:phi_ac} by
\begin{linenomath*}
\begin{alignat}{2}
  &\Phi^\text{new}_\text{au}(\tau) &&= a \left ( \frac{\bar{x}_c}{x_*} \right )^b
      + \Phi_\text{au}^\text{KCE}(\tau) \\
  &\Phi^\text{new}_\text{ac}(\tau) &&\equiv 1
\end{alignat}
\end{linenomath*}
with $a=0.1$ and $b=4$. The second term in the autoconversion equation
is the fit to the KCE data introduced in the previous section. The
first term replaces the constant contribution of the SB2001
formulation by a much smaller correction that increases rapidly with
the mean cloud droplet mass. This modification improves the agreement
with the KCE trajectories while preserving the physically desirable
$L_c$ scaling inherited from SB2001. The coefficient $a$ was chosen
to match the magnitude of the plateau inferred from the UODE results
in Fig.~\ref{Fig:phiau}b, while $b=4$ was selected empirically based on
the resulting timing errors and trajectory losses. The results are
relatively insensitive to the precise value of $b$. With this, the new
model equations read:
\begin{linenomath*}
\begin{alignat}{2}
  &AU_\text{new} &&= \frac{k_{cc}}{20 x_*} \, \frac{(\nu+2)(\nu+4)}{(\nu+1)^2} \,
                     L_c^2 \, \bar{x}_c^2 \, \Phi^\text{new}_\text{au}(\tau) \\
  &AC_\text{new} &&= k_{cr} \, L_c L_r 
\end{alignat}
\end{linenomath*}
The new parameterization gives similar results as SB2001 for $\thalf$
in Table 1, and improves over SB2001 in the log-loss and in the quadratic
loss.  The UODE still achieves the smallest logarithmic trajectory
loss, as expected since it was optimized using this metric.  The time
series of Figure~\ref{Fig:timeseries_states} hardly shows any differences between SB2001 and the
new formulation. The process rates, on the other hand, reveal that the
autoconversion rate is much closer to the KCE and UODE for the
reformulated parameterization. This shows that
$\Phi^\text{new}_\text{au}(\tau)$ is more consistent with the KCE than
SB2001. Interestingly, all three closures overestimate accretion for
small $\tau$. The fact that the new formulation does not suppress
accretion is hardly visible.
This indicates that the suppression of accretion in SB2001 primarily
compensates for the excessive autoconversion predicted at small
$\tau$, rather than representing an independent physical effect.

Due to the overestimation of autoconversion and the suppression of
accretion in the earliest stage of rain formation, SB2001
overestimates the number density of raindrops in this regime (Figure
\ref{Fig:timeseries_states}d). This is improved with the new
parameterization. In a mixed-phase cloud this can, for example, lead
to a more rapid glaciation with the new parameterization because
raindrops are larger making them more likely to freeze. The new
formulation may be particularly attractive in situations where
raindrops fall through a cloud layer from above. In such cases, SB2001
suppresses accretion for very small rainwater contents, potentially
delaying the onset of rain growth. Whether this has a measurable
impact in realistic three-dimensional simulations remains to be
seen. Indeed, SB2001 already included a validation of their
parameterization in one-dimensional simulations rather than idealized
box-model simulations and find a good agreement, suggesting that this
effect may be of limited practical importance. Given the wide use of
the SB2001 scheme, assessing the impact of the revised formulation in
LES and NWP models may nevertheless present an interesting topic for
future work.

\begin{table}[t]
\centering
\caption{Comparison of the KCE reference solution, SB2001, and the new
  parameterization for $L_0=0.5$\,g\,m$^{-3}$, different initial mean
  cloud droplet radii $\bar{r}_0$, and different shape parameters $\nu$. The
  first block shows the characteristic time $t_{50}$ (min), defined as
  the time at which 50$\,$\% of the initial cloud water has been converted
  to rain. The second block shows the logarithmic trajectory loss
  $\mathcal{J}$ relative to the KCE solution. Italics indicate
  the parameterization closest to KCE for $t_{50}$ and the lower
  trajectory loss for $\mathcal{J}$.}
\label{tab:comparison_nu}
\small
\begin{tabular*}{12cm}{@{\extracolsep{\fill}}c|rrr|rrr|rrr}
\toprule
\multicolumn{10}{c}{Characteristic time $t_{50}$ (min)}\\
\midrule
& \multicolumn{3}{c|}{$\nu=0$}
& \multicolumn{3}{c|}{$\nu=1$}
& \multicolumn{3}{c}{$\nu=2$}\\
$\bar{r}_0$ &
KCE & SB2001 & New &
KCE & SB2001 & New &
KCE & SB2001 & New\\
\midrule
11 &55.2&\emph{54.4}&58.1&77.3&\emph{77.5}&71.0&92.7&\emph{91.6}&76.3\\
12 &43.2&\emph{42.9}&47.3&61.8&\emph{60.4}&59.7&69.1&\emph{70.7}&65.0\\
13 &35.4&\emph{34.6}&38.5&49.4&48.4&\emph{49.9}&58.4&\emph{56.4}&55.0\\
14 &29.7&\emph{28.1}&31.3&40.0&\emph{39.6}&41.6&47.8&\emph{46.1}&46.4\\
15 &24.7&23.1&\emph{25.6}&33.9&32.8&\emph{34.6}&40.0&\emph{38.2}&39.0\\
16 &21.3&19.0&\emph{21.0}&29.1&27.4&\emph{28.8}&33.6&32.0&\emph{32.8}\\
17 &18.4&15.6&\emph{17.2}&24.8&23.0&\emph{24.1}&29.0&27.0&\emph{27.5}\\
18 &16.1&12.9&\emph{14.1}&21.2&19.3&\emph{20.1}&25.6&22.9&\emph{23.1}\\
19 &13.9&10.6&\emph{11.5}&19.2&16.3&\emph{16.7}&22.3&19.4&19.4\\
20 &12.3&8.7&\emph{9.4}&17.7&13.7&\emph{13.9}&20.0&\emph{16.4}&16.2\\
\end{tabular*}
\begin{tabular*}{12cm}{@{\extracolsep{\fill}}c|rr@{\hspace{15pt}}|rr@{\hspace{15pt}}|rr@{\hspace{20pt}}}
\toprule
\multicolumn{7}{c}{Logarithmic trajectory loss $\mathcal{J}$}\\
\midrule
& \multicolumn{2}{c|}{$\nu=0$}
& \multicolumn{2}{c|}{$\nu=1$}
& \multicolumn{2}{c}{$\nu=2$}\\
$\bar{r}_0$ &
SB2001 & New &
SB2001 & New &
SB2001 & New\\
\midrule
11&4.950&\emph{0.635}&6.981&\emph{1.137}&7.712&\emph{4.919}\\
12&3.831&\emph{0.905}&6.229&\emph{0.722}&4.423&\emph{0.426}\\
13&2.014&\emph{0.705}&4.822&\emph{0.667}&5.453&\emph{1.103}\\
14&1.566&\emph{0.304}&4.077&\emph{0.897}&5.278&\emph{1.510}\\
15&0.554&\emph{0.139}&2.609&\emph{0.637}&5.381&\emph{2.297}\\
16&0.234&\emph{0.046}&1.984&\emph{0.732}&2.369&\emph{1.028}\\
17&0.175&\emph{0.163}&1.498&\emph{0.862}&2.932&\emph{2.111}\\
18&0.367&\emph{0.298}&0.628&\emph{0.519}&2.236&\emph{2.189}\\
19&0.588&\emph{0.405}&\emph{0.331}&0.478&\emph{1.124}&1.592\\
20&0.894&\emph{0.563}&\emph{0.403}&0.637&\emph{1.061}&2.016\\
\bottomrule
\end{tabular*}
\end{table}

To validate the new scheme over a broader range of initial conditions,
Table~\ref{tab:comparison_nu} shows a comparison of both
parameterizations for $\bar{r}_0$ from 11-20 $\mu$m and for different
shape parameters $\nu$ of the initial Gamma distributions.  For the
characteristic time scale $t_{50}$, SB2001 generally agrees better
with the KCE solution for small $\bar{r}_0$, particularly for larger
values of $\nu$. In contrast, the new parameterization tends to
reproduce the KCE time scales more closely at intermediate and large
droplet radii, where the difference between the two schemes becomes
small. For large radii both parameterizations tend to underestimate the rain
formation time scale.

The trajectory losses reveal a different aspect of the
comparison. Despite the improved agreement of SB2001 in some cases for
$t_{50}$, the new parameterization generally achieves lower
logarithmic trajectory errors over a wide range of $\bar{r}_0$ and
$\nu$. This confirms again that SB2001 is optimized for the timing,
but less so for quantitative agreement over the full trajectory,
especially not for low $L_r$. The remaining discrepancies of the new
parameterization are mainly confined to large $\nu$ and small $\bar{r}_0$,
where SB2001 retains an clear advantage in onset time, but not in
trajectory error. Overall, the results demonstrate that the new
formulation provides an improved representation of the evolution of
rain formation, while accepting a modest loss in onset-time accuracy
in a limited region of the parameter space.

\section{Summary and Conclusions}

The warm-rain parameterization problem can be formulated as the search
for a finite-dimensional system of ordinary differential equations that
approximates the evolution of the kinetic collection equation (KCE).
The main difficulty is the closure of the highly nonlinear
autoconversion process. Existing analytical parameterizations therefore
rely on assumptions whose impact and validity are often difficult to
assess.

In this study, the framework of universal ordinary differential
equations (UODEs) has been applied to learn the warm-rain closure
directly from KCE trajectories. Unlike most previous machine learning
approaches, the UODE is trained only on time series of the bulk state
variables and does not require process rates as training targets. The
resulting model replaces the autoconversion and accretion tendencies,
while the remaining self-collection terms of the SB2001 parameterization
are retained.

The learned UODE reproduces the KCE trajectories with high accuracy over
the range covered by the training data and substantially improves upon
the SB2001 parameterization in terms of the logarithmic trajectory loss.
Although the decomposition into autoconversion and accretion was not
prescribed during training, the learned closure yields process rates
that closely resemble those diagnosed independently from the KCE.
This demonstrates that the four bulk state variables contain sufficient
information to infer a physically meaningful approximation of the
underlying warm-rain operator. At the same time, the reduced skill of
the UODE outside the range of the training data shows that the learned
representation does not fully recover the underlying physical scaling
of the KCE, but instead provides an accurate approximation within the
sampled parameter space.

Perhaps the most important result of this study is that the learned
UODE provides insight into the structure of analytical warm-rain
parameterizations. The results suggest that the suppression of
accretion in SB2001 primarily compensates for an overestimation of
autoconversion at small rain fractions. Guided by this observation, a
reformulation of the SB2001 closure is proposed in which the
autoconversion correction is modified and the empirical accretion
correction is removed. The resulting analytical parameterization
retains the simplicity of SB2001 while providing a closer approximation
to the KCE and generally reducing both logarithmic and quadratic
trajectory errors. The improvement is not uniform across all initial
conditions and metrics, however: SB2001 remains more accurate for the
characteristic conversion time in parts of the tested parameter space.

The present work therefore demonstrates that scientific machine
learning can be used not only to construct surrogate models, but also
as a tool for analyzing and improving existing physical
parameterizations. Rather than replacing analytical parameterizations,
the UODE serves as a flexible reference model that can reveal
deficiencies in existing closures and guide the development of
improved analytical formulations.

Several limitations remain. The UODE training and evaluation consider
only collision and coalescence in an idealized box model, use a single
collection kernel, and are restricted to the shape parameter $\nu=1$.
The broader validation across different shape parameters is performed
only for the proposed analytical parameterization and SB2001. Moreover,
the closure problem itself remains ill-posed for two-moment bulk
schemes, so no deterministic closure based on the retained variables
can be expected to reproduce the KCE exactly. Future work should
therefore investigate whether the proposed analytical reformulation
improves simulations in large-eddy, numerical weather prediction, and
climate models, and whether similar UODE approaches can be applied to
more comprehensive cloud microphysics schemes.

An interesting direction for future work is a more accurate
representation of the cloud and rain droplet number equations within
the bulk framework, particularly when additional processes such as
sedimentation and evaporation are included. The UODE approach also
appears promising for analyzing higher-order closures in
single-category warm-rain schemes.

Overall, this study illustrates how scientific machine learning can
serve not only as a means of emulating physical models, but also as a
tool for extracting their underlying structure and guiding the
development of simpler and more physically consistent
parameterizations.


\newpage

\section*{Code and Data Availability}

The training and testing data, Python and Julia codes, and neural
network coefficients of the UODE are available from Zenodo
\url{https://doi.org/10.5281/zenodo.22201757} and the public
repository \url{https://gitlab.com/axelseifert/warmrainuode}. The
Lagrangian microphysics model McSnow used to generate the KCE training
and testing data is part of the ICON modeling framework and is
available from the public repository
\url{https://gitlab.dkrz.de/mcsnow/mcsnow}.

\section*{Acknowledgments}

The author thanks Roland Wirth, Dmitrii Mironov, and Christoph Siewert
for helpful comments on the manuscript.  ChatGPT (OpenAI) was used for
assistance with software development, code debugging, and manuscript
editing.

\bibliographystyle{copernicus}
\bibliography{uode.bib}

\end{document}